\documentclass[fleqn,10pt, twocolumn]{wlscirep}
\usepackage{setspace}
\usepackage{textcomp}

\usepackage{braket}
\usepackage{siunitx}
\usepackage[position=top]{subfig}
\title{Resolving the temporal evolution of line broadening in single quantum emitters}

\author[1,*]{Christian Schimpf}
\author[1]{Marcus Reindl}
\author[2,3,4]{Petr Klenovsk\'y}
\author[1]{Thomas Fromherz}
\author[1]{Saimon F. Covre Da Silva}
\author[1]{Julian Hofer}
\author[5]{Christian Schneider}
\author[5,6]{Sven Höfling}
\author[7]{Rinaldo Trotta}
\author[1]{Armando Rastelli}

\affil[1]{Institute of Semiconductor and Solid State Physics, Johannes Kepler University, Linz 4040, Austria}
\affil[2]{Department of Condensed Matter Physics, Faculty of Science, Masaryk University, Kotl\'a\v{r}sk\'a~267/2, 61137~Brno,Czech~Republic}
\affil[3]{Central European Institute of Technology, Masaryk University, Kamenice 753/5, 62500~Brno, Czech~Republic}
\affil[4]{Czech Metrology Institute, Okru\v{z}n\'i 31,63800~Brno, Czech~Republic}
\affil[5]{Technische Physik and Wilhelm Conrad Röntgen Research Center for Complex Material Systems, Universität Würzburg, 97074 Würzburg, Germany}
\affil[6]{SUPA, School of Physics and Astronomy, University of St Andrews, St Andrews, KY16 9SS, UK}
\affil[7]{Department of Physics, Sapienza University, 00185 Rome, Italy}

\affil[*]{Corresponding author: christian.schimpf@jku.at}

\begin{abstract}
Light emission from solid-state quantum emitters is inherently prone to environmental decoherence, which results in a line broadening and in the deterioration of photon indistinguishability. Here we employ photon correlation Fourier spectroscopy (PCFS) to study the temporal evolution of such a broadening in two prominent systems: GaAs and In(Ga)As quantum dots. Differently from previous experiments, the emitters are driven with short laser pulses as required for the generation of high-purity single photons, the time scales we probe range from a few nanoseconds to milliseconds and, simultaneously, the spectral resolution we achieve can be as small as $\sim\SI{2}{\micro eV}$. We find pronounced differences in the temporal evolution of different optical transition lines, which we attribute to differences in their homogeneous linewidth and sensitivity to charge noise. We analyze the effect of irradiation with additional white light, which reduces blinking at the cost of enhanced charge noise. Due to its robustness against experimental imperfections and its high temporal resolution and bandwidth, PCFS outperforms established spectroscopy techniques, such as Michelson interferometry. We discuss its practical implementation and the possibility to use it to estimate the indistinguishability of consecutively emitted single photons for applications in quantum communication and photonic-based quantum information processing.
\end{abstract}

\begin{document}
\maketitle

\section{Introduction}
Semiconductor quantum dots (QDs) are excellent resources of non-classical light for modern photonics \cite{Senellart2017}. They enable pivotal single photon applications in the field of quantum communication and quantum information processing, such as quantum teleportation~\cite{Reindl2018} and Boson sampling~\cite{Tillmann2013, Wang2018}. A key parameter in these applications is the \textit{indistinguishability} of consecutively emitted photons, which is usually quantified as the visibility of the Hong-Ou-Mandel (HOM) interference between two photons impinging on a 50:50 beam splitter~\cite{Mandel1987}\cite{Legero2003}. In turn, this directly impacts the error rate in envisioned applications. The photon indistinguishability of solid state light emitters is predominantly reduced by coupling to the phonon bath~\cite{Reigue2017} and by a fluctuating charge environment \cite{Kuhlmann2013}, both leading to a dephasing between the single photon wave packets \cite{Kambs2018}. In semiconductors, charge noise can stem from ionisation of impurities or from photogenerated carriers, which produce shifts of the central emission energy of a nearby QD due to the quantum confined Stark effect \cite{Miller1984}. This results in inhomogeneously broadened spectral lines, usually observed in time-averaged micro-photo-luminescence ($\mu$PL) spectroscopy. The 1/f noise frequency spectrum determines the indistinguishability at different timescales, which can be important up to hundreds of nanoseconds, as in multiphoton experiments relying on time-to-path demultiplexing~\cite{lenzini2017active, Wang2018}.

The temporal dynamics of the dephasing mechanisms usually elude established spectroscopy methods, such as $\mu$PL and Michelson interferometry \cite{Santori2002, Berthelot2006}, due to relatively long integration times required. Scanning resonance fluorescence techniques have been used to probe timescales down to the microseconds range \cite{Kuhlmann2015}, shedding light on noise sources, but cutting off crucial parts of the frequency spectrum. In addition, this technique is based on excitation via a continuous wave laser, while true single-photon generation requires excitation pulses with temporal duration substantially shorter than the excitonic lifetime. The most reliable way of quantifying the indistinguishability so far is a direct measurement of the interference visibility in a HOM experiment, using varying delay times to probe photons from different excitation cycles \cite{Loredo2016, Wang2016, Thoma2016}. However, this approach has limited temporal resolution due to the inefficient way of varying the delay time and the maximum achievable delay is limited by optical fiber losses.

Photon Correlation Fourier Spectroscopy (PCFS), first introduced by Brokmann et al, \cite{Brokmann2006} can provide quantitative information about the emission line broadening  at time scales down to the source lifetime (usually sub-nanoseconds) and up to several hours, all in one measurement. It therefore allows an estimation of the trend of the HOM visibility with increasing photon delay. As a Fourier method, it benefits from a high energy resolution compared to diffraction spectroscopy, similar to Michelson interferometry, but without the high requirements in mechanical stability of the optical components. We also stress that, different from scanning resonance fluorescence, PCFS is completely independent on the excitation method (optical or also electrical) and can thus be used to characterized the emitter properties "in operando", i.e. under exactly the same conditions used to generate  single photons. 

In this article we employ PCFS to investigate line broadening in two QD systems: GaAs QDs and In(Ga)As QDs. In(Ga)As QDs obtained by the Stranski-Krastanow growth method have been used over the last two decades to preform several pioneering experiments \cite{Michler2000,Santori2002, Akopian2006, Yuan2002, Senellart2017, Loredo2016}. GaAs obtained via the local droplet etching method~\cite{Heyn2010, Huo2013} have emerged more recently \cite{Gurioli2019}. They exhibit intrinsically short lifetimes, a near-zero multi-photon emission probability \cite{Schweickert2018}, and a high indistinguishability of consecutively emitted photons \cite{Huber2016} \cite{Reindl2018Arxiv}, despite the non-lifetime-limited time-averaged linewidth \cite{Reindl2017}. In addition, these QDs can be used to generate polarization-entangled photon pairs with near-unity entanglement fidelity using the biexciton-exciton recombination cascade \cite{Huber2018}. These properties recently allowed performing entanglement swapping between subsequently emitted photon pairs \cite{BassoBasset2019Arxiv}, where the success rate is mostly determined by the indistinguishability of the photons used for the Bell state measurement.\\
We therefore investigate line broadening for the emission lines XX and X corresponding to the biexciton-to-exciton and the exciton-to-groundstate decays for both material systems. We point out differences and common properties based on the information gained from the measurements.

\section{Methods}

\begin{figure}[htbp!]
    \centering\includegraphics[trim={5mm 0 0 0},width=9cm]{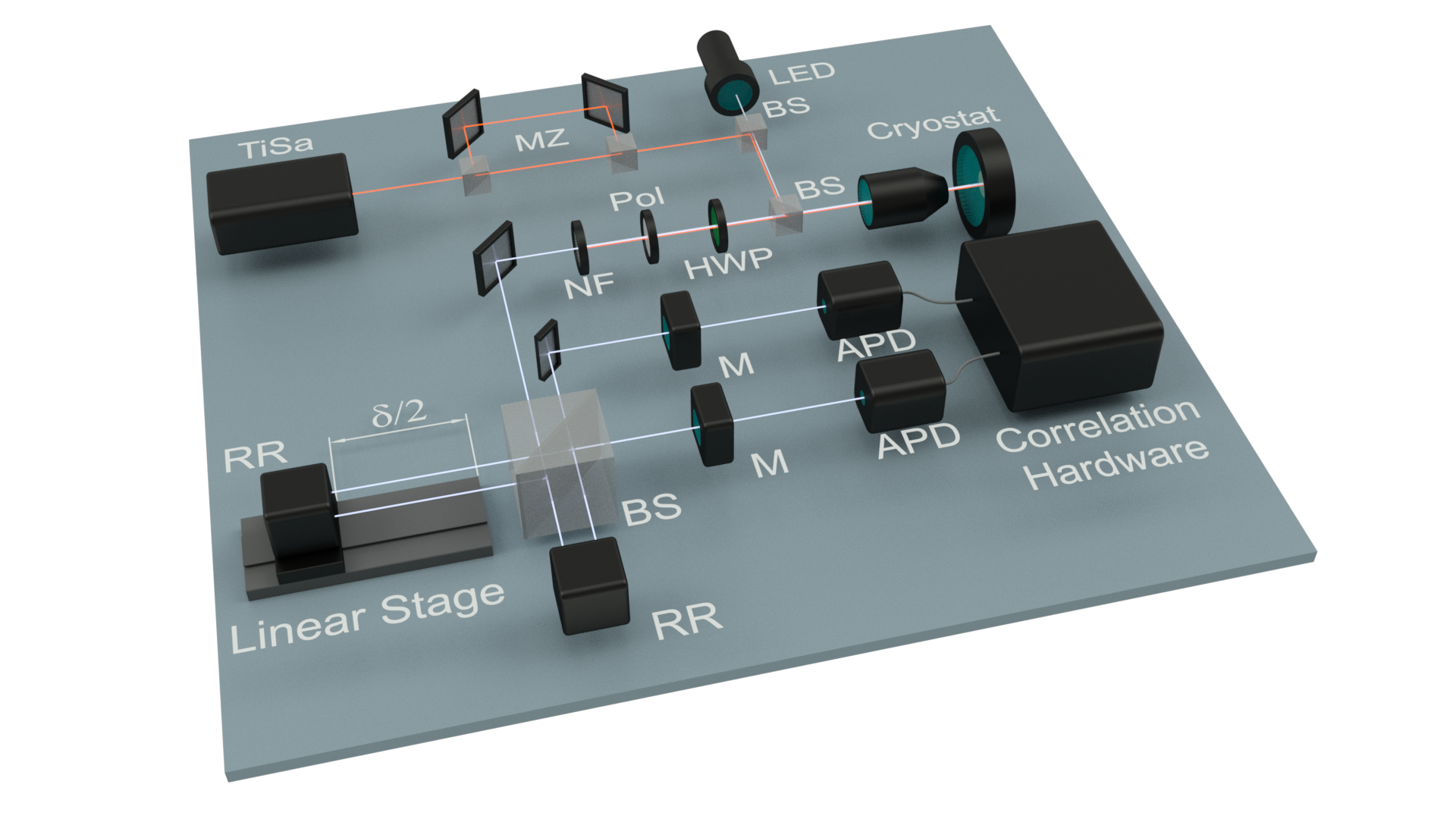}
\caption{Sketch of the experimental setup. The quantum dots (QDs) are placed in a cryostat and excited by a pulsed TiSa laser with a pulse duration of about \SI{5}{ps} and a separation of \SI{12.5}{ns}. Each pulse is doubled by an unbalanced Mach-Zehnder interferometer (MZ) with a time delay of \SI{2}{ns}. Additional white light provided by an LED can be coupled into the excitation path. The emitted photons (XX and X) from the QD's decay cascade can be filtered in polarization by a half-wave-plate (HWP) and a polariser (Pol), so that only one fine structure component is selected. Variable notch filters (NF) with a bandwidth of \SI{0.4}{nm} reject the stray light from the laser. The signal is guided to a Michelson interferometer, consisting of two retroreflectors (RR), one of them on a motorized linear stage with \SI{300}{mm} travel range. The reflected beams interfere at a beam splitter (BS). Both beams are spectrally filtered by monochromators (M) to select either the X or XX emission line and finally measured by avalanche photodiodes (APDs) - connected to correlation hardware.}
\label{fig:setup}
\end{figure}
\begin{figure}[htbp!]
\centering
\includegraphics[width=90mm]{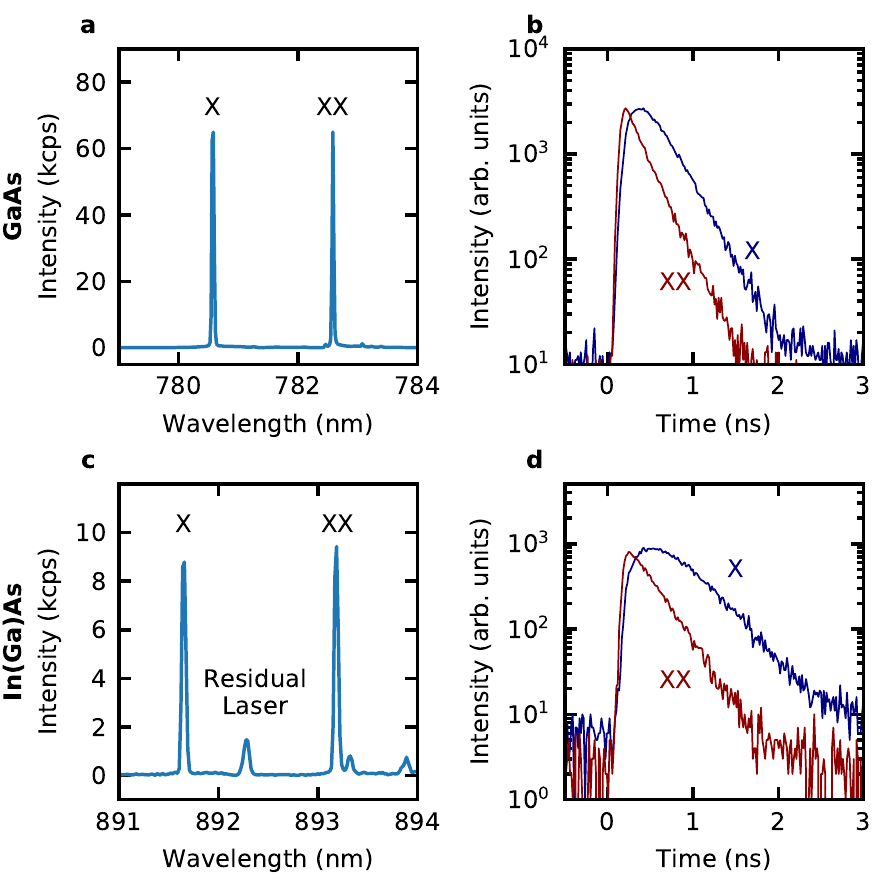}
\caption{(a) Emission spectra of a GaAs quantum dot (QD) under resonant two-photon excitation (TPE) - measured by micro photoluminescence ($\mu PL$) spectroscopy. XX and X mark the emission lines of the biexcition to exciton- and the excition to groundstate radiative transitions, respectively. (b) Time traces of the X and XX emission from the GaAs QD. Convoluted fits, considering the instrument response function, yield lifetimes of $T_{1_,\text{XX}}=\SI{115(4)}{ps}$ and $T_{1_,\text{X}}=\SI{267(14)}{ps}$. (c) Emission spectra of an In(Ga)As QD under TPE. (d) Time traces from the In(Ga)As QD. The fitted lifetimes are $T_{1_,\text{XX}}=\SI{186(6)}{ps}$ and $T_{1_,\text{X}}=\SI{351(15)}{ps}$.}
\label{fig:spectra}
\end{figure}

The GaAs QDs studied here were grown with the local droplet etching method by molecular beam epitaxy (MBE) in an AlGaAs matrix and embedded in a simple planar cavity. The sample structure is similar to that used in recent publications~\cite{Schweickert2018,Huber2018,Reindl2018,Reindl2018Arxiv,BassoBasset2019Arxiv}.
The In(Ga)As QDs under investigation were also fabricated by MBE, using the Stranski-Krastanow method. The low density QD layer is embedded in a $\lambda$/n thick GaAs layer, surrounded by 24 (5) AlGaAs/GaAs mirror pairs in the bottom (top) DBR (details can be found in \cite{Maier2014}). Buried nanohills, which naturally occur during the eptaxial growth, provide lateral confinement to the optical mode, and thus strongly enhance the photon extraction efficiency \cite{Maier2014}, while the high optical emitter quality is reflected by high degrees of photon indistinguishability of consecutively emitted photons \cite{He2013,Gerhardt2018}.\\
Figure~\ref{fig:setup} illustrates the experimental arrangement for PCFS. The QDs are cooled to \SI{5}{K} in a He flow cryostat and excited by a pulsed TiSa laser with a pulse duration of about \SI{5}{ps} and a repetition period of \SI{12.5}{ns}to produce trains of photon pairs, similar to recent experiments \cite{Schweickert2018, Liu2019, Muller2014, Chen2018}.
 To access both the XX and X lines, we employ a two-photon excitation scheme (TPE) \cite{Jayakumar2013,Muller2014}, with the laser energy tuned precisely to the half of the biexciton energy. The resulting spectra are shown in figure~\ref{fig:spectra} (a) and (c), such as the respective decay time traces in (b) and (c). To decrease the total acquisition time and to enhance the temporal resolution, we double the average repetition rate of the laser by an unbalanced Mach-Zehnder interferometer (MZ), yielding two pulses with a time delay of \SI{2}{ns} per laser pulse.
In addition to the laser, white light from an LED with variable intensity can be coupled into the same excitation path.

The spectral separation between X and XX lines - typically about \SI{2}{nm} (\SI{4}{meV}) for GaAs QDs and \SI{1.5}{nm} (\SI{3}{meV}) for the here employed In(Ga)As QD - allows a selection of the lines via standard diffraction monochromators (M) before detection, see figure~\ref{fig:setup}. The scattered laser light, located between the X and XX lines, is reflected by tunable notch filters (NF) with a bandwidth of \SI{0.4}{nm}. 
Since the exciton energy level is split by a fine structure splitting (FSS), the X and XX lines actually consist of doublets with orthogonal linear polarization~\cite{Bayer2002}. In the setup  we can select one fine structure component of each doublet by rotating the polarization plane of the emission signal by a half wave plate (HWP), until one component is cancelled by a subsequent polarizer.

The interferometer of the PCFS setup consists of a 50:50 beam splitter (BS) and two retroreflectors (RR), where one RR is mounted on a motorized linear stage. The position of the stage defines the relative optical path difference $\delta$ between the two arms of the interferometer. In contrast to standard Michelson interferometry, PCFS detects the signal of \textit{both} BS outputs via two avalanche photodiodes (APDs), which are connected to correlation hardware. In our experiment we use APDs with a time resolution of \SI{500}{ps}.

\begin{figure}[htbp!]
\centering
\includegraphics[width=90mm]{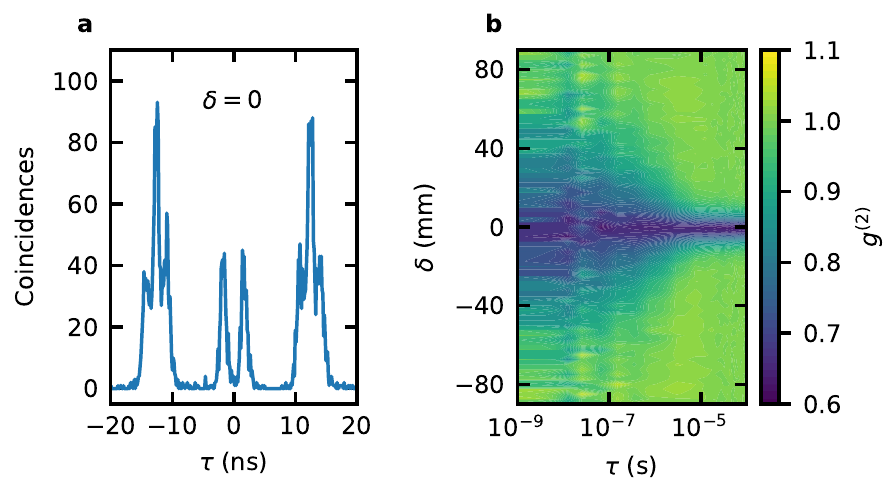}
\caption{(a) Recorded coincidence histogram (corresponding to the unnormalized $g^{(2)}(\tau)$) of the X emission of a GaAs QD for a time delay $\tau$ at the nanosecond timescale and fixed optical path delay of $\delta=0$. The absence of a peak at $\tau=0$ reflects the single photon emission characteristics of the QD. (b) Normalized second order correlation function $g^{(2)}$ of the X photons as a function of $\delta$ and $\tau$. The value (ideally) drops to $0.5$ at $\delta=0$ and converges to $1$ at sufficiently large values of $\delta$, with a functional behaviour depending on the dephasing mechanisms at different $\tau$.}
\label{fig:g2}
\end{figure}

The PCFS measurement is performed by positioning the linear stage at several equidistant positions, yielding a discrete set of $N$ different optical path differences $\{\delta_i\}$, $i\in\{0,\dots,N-1\}$, with a constant spacing of $\Delta=\delta_{i+1}-\delta_{i}$. In Michelson interferometry, usually the intensity contrast (or "visibility") between the interference maxima and minima at one BS output is determined. In PCFS the interference fringes are averaged by slowly moving the linear stage with a constant velocity $v$, i.e. no high precision linear stage is needed  (in contrast to Michelson interferometry).

To illustrate the working principle of PCFS we first consider a monochromatic plane wave with a time-dependent angular frequency $\omega(t)$. In the classical picture, the resulting intensities at the respective BS outputs A and B are then proportional to

\begin{equation}
\begin{aligned}
I_A(t,\delta_i) &\propto 1+\text{cos}\big[ (2vt+\delta_i)\,\omega(t)/c \big],\\
I_B(t,\delta_i) &\propto 1-\text{cos}\big[ (2vt+\delta_i)\,\omega(t)/c \big].
\end{aligned}
\label{eq:int}
\end{equation}

In this picture, the frequency of the emitted light would be well defined at any time. We know from Heisenberg's principle that this is impossible for light stemming from spontaneous emission and that the variable $\omega$ has to follow a probability distribution with a finite natural linewidth, depending on the lifetime $T_1$ of the emitter. Dephasing processes lead to further broadening of this distribution. 

Fluctuations of $\omega$ in semiconductor structures usually occur on timescales much smaller than the measurement times of commonly-used spectroscopy methods. Those techniques only reflect the time average of the frequency distribution, concealing its temporal evolution. PCFS allows us to recover the time dependency by exploiting the time resolution of our single photon detectors (APDs) and determining the second order correlation function

\begin{equation}
    g^{(2)}(\tau,\delta_i)=\frac{\braket{I_A(t,\delta_i)I_B(t+\tau,\delta_i)}_t}{\braket{I_A(t,\delta_i)}_t \braket{I_B(t+\tau,\delta_i)}_t},
\label{eq:g2}
\end{equation}

\noindent
where $\braket{\dots}_t$ denotes the time average over the integration time $T$ and $\tau$ is the time difference between detection events (see later \eqref{eq:g2_av}).

To avoid confusion, we stress at this point that the optical path difference $\delta$ gives us access to the spectral distribution, similar to conventional Michelson interferometry. It is in fact evident from equation~\ref{eq:int} that interference will produce a drop in the coincidence counts between the two detectors and that a relatively low value of $g^{(2)}(\tau,\delta_i)$ is thus expected as long as $\delta_i$ is small compared to the coherence length of the source, i.e., as long as interference fringes are detectable. The high temporal resolution and large temporal bandwidth on the determination of $\tau$ adds the desired temporal information.

The velocity $v$ in PCFS is particularly relevant and should satisfy the following conditions: 1) The travelling distance of the stage within $T$ has to be small, so that $2vT \ll \Delta$. This condition is equivalent to asserting that the interference visibility has a well defined value for a given value of path difference $\delta_i$.  2) The travelling distance must be sufficiently long in order to average over several interference fringes, so that $2vT \gg 2\pi c / \omega_0$, where $\omega_0$ is the mean frequency of the emitted signal. 3) The change in $\delta$ within the maximum time delay of interest $\tau_{max}$ has to be substantially lower than the coherence length of the emitter, i.e. $2v\tau_{max} \ll cT_2$ (see \cite{Brokmann2006} for details).

Figure \ref{fig:g2} (a) shows the unnormalized $g^{(2)}(\tau)$ of the X signal of a GaAs QD for a fixed $\delta=0$. The triple-peak pattern originates from the pulse sequence of the excitation laser, doubled by the MZ. The peak at $\tau=0$ is absent due to the single photon emission characteristics of the QD \cite{Schweickert2018}. The single photons arrive within a minimum time delay of $\SI{2}{ns}$ at the two detectors, yielding the two peaks around $\tau=0$ and thus defining the time resolution $\tau_{min}=\SI{2}{ns}$. This threshold solely arises from the specific measurement arrangement and can, in principle, be pushed down to the emitter lifetime $T_1$ by further reducing the delay of the MZ or by utilizing a CW pump laser. The next value of $\tau$ accessible with the present configuration corresponds to the delay between two subsequent laser pulses, i.e. $\SI{12.5}{ns}$.

We evaluate the $g^{(2)}$ for time delays up to $\tau_{max}=\SI{1}{ms}$, which turned out to be sufficiently high to cover the slowest frequency fluctuations in the used QDs. The total observable noise frequency range in this case is thus from \SI{1}{kHz} to \SI{500}{MHz}, which extends well beyond \SI{100}{kHz}, previously accessed by resonant fluorescence \cite{Kuhlmann2015}. After repeating the $g^{(2)}$ measurement for all different $\delta_i$, a two-dimensional map of $g^{(2)}$ values as a function of $\tau$ and $\delta$ can be constructed, as shown in figure \ref{fig:g2} (b). Provided that the aforementioned conditions regarding $v$ are fulfilled, the value of $g^{(2)}$ always drops to (ideally) $0.5$ at $\delta=0$. The measured value of about $0.6$ is a result of experimental imperfections, such as a non-perfect overlap of the interfering beams at the BS. For $\delta \neq 0$ and fixed $\tau$ the $g^{(2)}$ value increases towards $1$ (when corrected for the "blinking" of the emitter, explained in section \ref{sec:white}), following a functional behaviour, which depends on the spectral broadening mechanisms acting within a time delay $\tau$: the stronger the broadening, the narrower the dip of $g^{(2)}$. From the 2D map we already qualitatively see that the spectral fluctuations of the emitter increase with increasing time delay $\tau$. The maximum path delay induced by the linear stage therefore has to be well above the maximum coherence length $2cT_{1}$ of the emitted signal, in order not to lose information.

To draw conclusions about the energy distribution of our signal, we combine \eqref{eq:int} and \eqref{eq:g2}:

\begin{equation}
g^{(2)}(\tau,\delta_i)=1-\frac{1}{2T} \int \limits_0^T \text{cos}(\zeta_\tau(t)\,\delta_i/c)\, dt, 
\label{eq:g2_av}
\end{equation}

\noindent
where $\zeta_\tau (t)= \delta\omega(t+\tau)-\delta\omega(t)$ describes a random frequency shift between $t$ and $t+\tau$, with $\delta\omega(t)$ the frequency deviation from $\omega_0$. For sufficiently long $T$, all possible frequency fluctuations will occur, hence we can apply the ergodic theory to substitute the time average with the ensemble average:

\begin{equation}
g^{(2)}(\tau,\delta_i)=1-\frac{1}{2} \int \limits_{-\infty}^{\infty} \text{cos}(\zeta\,\delta_i /c)p_\tau(\zeta)\,d\zeta,
\label{eq:g2_zeta}
\end{equation}

\noindent
where the normalized probability distribution $p_\tau(\zeta)$ over the random variable $\zeta=\zeta_\tau(t=0)$ contains the desired information about the time-dependent spectral properties of the measured signal. We identify the integral in \eqref{eq:g2_zeta} as Fourier transform \cite{Brokmann2006}. Applying the discrete inverse Fourier transform yields the values for $p_\tau(\zeta)$,

\begin{equation}
    p_\tau(\zeta_j) = 2 \biggl| \sum_{k=0}^{N-1} \left[ 1- g^{(2)}_{\tau}(\delta_k) \right]e^{-\frac{2\pi i}{N}\,k\,j} \biggl| ^2,
\end{equation}

\noindent
with $j\in \{0,\dots,N-1\}$ and $\zeta_j=2\pi cj/N\Delta$. The value of $g^{(2)}_{\tau}(\delta_k)$ is evaluated by integrating over an appropriate time bin $\Delta \tau$ (see supplementary).

The maximum induced optical path difference $N\Delta=\delta_{N-1}-\delta_0$ defines the frequency resolution by $\zeta_{min}=2\pi c/N\Delta$. For our linear stage with a maximum travel range of $\SI{300}{mm}$, the energy resolution results in $\epsilon_{min}=\hbar \zeta_{min} = \SI{2.1}{\micro eV}$. The measurable spectral range is given by $\zeta_{max}=2\pi c/\Delta$. For $\Delta=\SI{2.5}{mm}$, the energy range then results in $\epsilon_{max}=\hbar \zeta_{max} = \SI{245}{\micro eV}$.

It should be noted that the distribution of spectral shifts $p_{\tau}(\zeta)$ can \textit{not} directly be identified as the spectral distribution of the measured light emission. To elaborate this statement, we take a closer look at the formal definition:

\begin{equation}
    p_\tau(\zeta) = \braket{ \int \limits_{-\infty}^{\infty} s_t(\omega)s_{t+\tau}(\omega+\zeta)\,d\omega }.
\label{eq:pzeta_s}
\end{equation}

In \eqref{eq:pzeta_s} the emission frequency $\omega$ is generalized to the time-dependent homogeneous distribution $s_t(\omega)$. In the case of QDs, it ideally corresponds to the natural Lorentzian spectral distribution with a linewidth of $\Gamma=1/T_1$, often referred to as the \textit{Fourier limit}. The integral in \eqref{eq:pzeta_s} represents the cross-correlation function between the spectral distributions at the time $t$ and at $t+\tau$, which could differ because of a spectral shift within $\tau$. The ensemble average ($\braket{\dots}$) over all possible realisations of frequency fluctuations leads to $p_\tau(\zeta)$.

For the limiting case of $\tau \rightarrow 0$ no frequency fluctuations have occurred and $s_{t+\tau}\rightarrow s_t$. Consequently, $p_{0}(\zeta)$ corresponds to the auto-correlation function of the Fourier-limited line shape, which is again a Lorentzian distribution, but with a FWHM of $2\Gamma$. For the second limiting case of $\tau \rightarrow \infty$, all possible frequency shifts have happened between $t$ and $t + \tau$. This situation corresponds to the regime accessed by standard Michelson interferometry and $p_{\infty}(\zeta)$ represents the auto-correlation function of the time averaged spectral distribution. In the case of a Gaussian broadening with a FWHM of $\Sigma$, the resulting observed $p(\zeta)$ is again a Gaussian distribution with a FWHM of $\sqrt{2}\Sigma$.

\begin{figure}[htbp!]
\centering
\includegraphics[width=90mm]{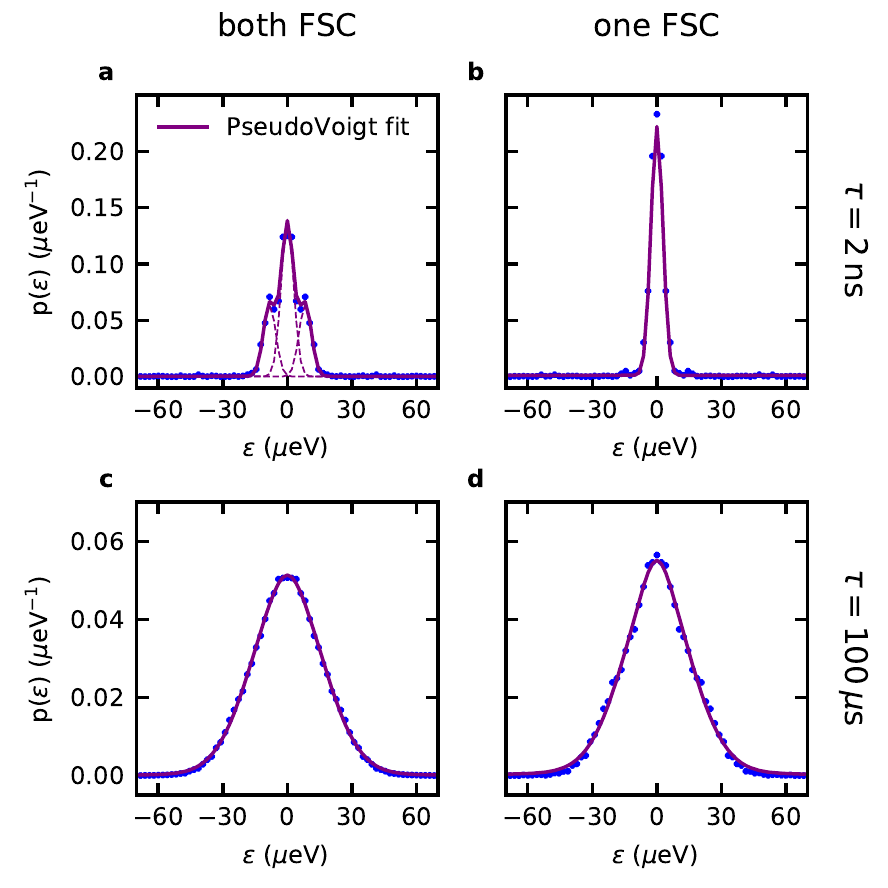}
\caption{(a) Distribution of the spectral shifts $p(\epsilon)$ at $\tau=\SI{2}{ns}$, for both fine structure components (FSCs) of the X signal of a GaAs QD. The doublet appears as a triplet in $p(\epsilon)$. The fitted sidepeaks are located at the energy corresponding to the fine structure splitting $S=\SI{8.3(1)}{\micro eV}$ and show a FWHM of \SI{6.81(11)}{\micro eV}. (b) One FSC of the same signal evaluated at $\tau=\SI{2}{ns}$ with a fitted FWHM of $\SI{6.48(8)}{\micro eV}$. (c,d) Distribution for the X emission at $\tau=\SI{100}{\micro s}$, for both and one FSC, respectively. The distinct peaks from the FSCs are obscured, the FWHM of the inhomogeneously broadened lines are \SI{39.2(2)}{\micro eV} and \SI{38.4(3)}{\micro eV}, respectively.}
\label{fig:FSS}
\end{figure}

\section{Measurements and results}

\subsection{Resolving the fine-structure of inhomogeneously broadened lines with PCFS} 
To illustrate the high spectral resolution of PCFS and gain further insight in the information provided by the technique, we focus here on the fine-structure-split X emission of a GaAs QD. 

Figure \ref{fig:FSS} shows an excerpt of a PCFS measurement, evaluated at two different time delays (a,b) $\tau=\SI{2}{ns}$  and (c,d) $\tau=\SI{100}{\micro s}$. Instead of using the angular frequency shift $\zeta$, we plot the results as a function of the energy shift $\epsilon=\hbar\zeta$. The panels on the left show measurements of both fine structure components (FSCs). These components represent a static doublet in the spectral distribution, separated by the fine structure splitting (FSS). The value of the FSS was determined from polarization-resolved photo-luminescence (PL) spectra as $S=\SI{8.2(2)}{\micro eV}$ (see supplementary for details). According to \eqref{eq:pzeta_s}, a static doublet of homogeneously broadened lines results in a peak triplet in $p(\epsilon)$, as seen in panel (a), corresponding to all different combination of spectral shifts: The peak at $\zeta=0$ represents the absence of a spectral shift (if two photons stem from the same FSC). The two identical side peaks correspond to an equally probable positive or negative energy shift of $\pm S$ (if two photons stem from different FSCs). The triplet is fitted with the sum of three Pseudo Voigt functions. The FWHM of each peak is $\SI{6.81(11)}{\micro eV}$ and the mean energies of the two side peaks yield $S=\SI{8.3(1)}{\micro eV}$, which is in good agreement with the value measured by $\mu$PL. Figure \ref{fig:FSS} (b) corresponds to the same X emission signal, when selecting only one FSC. In that case, $p(\epsilon)$ shows only one line with fitted FWHM of \SI{6.48(8)}{\micro eV}. The expected FWHM of a Fourier limited Lorentzian spectral line is $2\Gamma_{\text{X}}=\SI{4.97(19)}{\micro eV}$. The discrepancy indicates the presence of a broadening mechanism, already active at $\tau=\SI{2}{ns}$, probably due to both interaction with lattice vibrations \cite{Reigue2017} and charge noise.

As previously stated when discussing figure \ref{fig:g2} (b), the spectral broadening is expected to increase with time delay $\tau$. In figures \ref{fig:FSS} (c) and (d) the same two measurement configurations (both and one FSC) are evaluated at $\tau=\SI{100}{\micro s}$. In these cases, the inhomogeneous broadening due to charge noise \cite{Kuhlmann2013} in the QD surroundings dominates the line shapes. The triplet shown in (a) is completely obscured and the fitted FWHM of (c) and (d) of \SI{39.2(2)}{\micro eV} and \SI{38.4(3)}{\micro eV} are found to be similar. A Michelson measurement of one FSC yields a linewidth of \SI{20.9(2)}{\micro eV}. We recall that $p(\epsilon)$ corresponds to an auto-correlation function of a spectral distribution (\eqref{eq:pzeta_s}). Therefore, the real linewidth can be deduced from $p(\zeta)$, if the original \textit{lineshape} is known. If not, we can still assume the Lorentzian and Gaussian distribution as limiting cases to estimate the original linewidth. Assuming a predominantly Lorentzian lineshape, the results of PCFS and Michelson interferometry are in sufficiently good agreement.

From the above discussion it is clear that PCFS allows to overcome inhomogenous broadening and thus to access spectrally close lines even without resorting to polarization selection. 

\subsection{Temporal evolution of the X and XX spectral line broadening from a GaAs QD and an In(Ga)As QD}

\begin{figure}[htbp]
\centering\includegraphics[width=90mm]{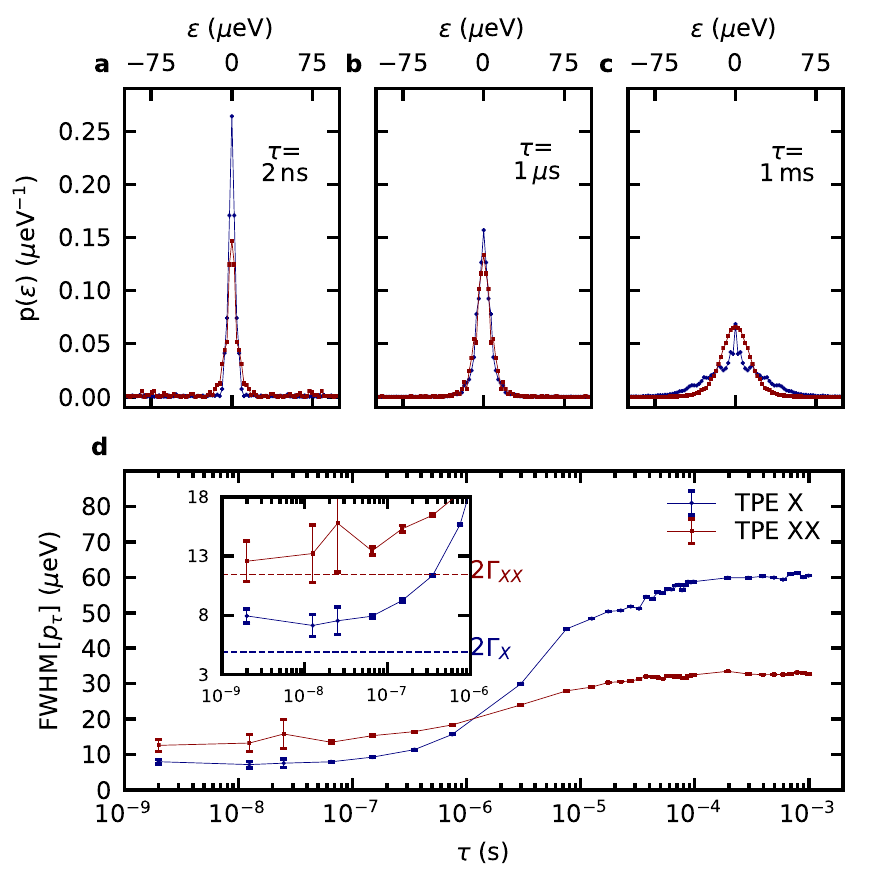}
\caption{(a-c) Distribution of spectral shifts $p(\epsilon)$ of the X and XX emission of an GaAs QD at time delays $\tau$. One FSC was selected by a polarizer. (d) Full width at half maximum (FWHM) of $p(\epsilon)$ as a function of $\tau$ in the range from \SI{2}{ns} to \SI{1}{ms}. The dashed lines in the inset correspond to the X and XX Fourier limits of $2\Gamma_{\text{X}}=\SI{4.93(25)}{\micro eV}$ and $2\Gamma_{\text{XX}}=\SI{11.45(40)}{\micro eV}$, respectively.}
\label{fig:XXevo}
\end{figure}

\begin{figure}[htbp]
\centering\includegraphics[width=90mm]{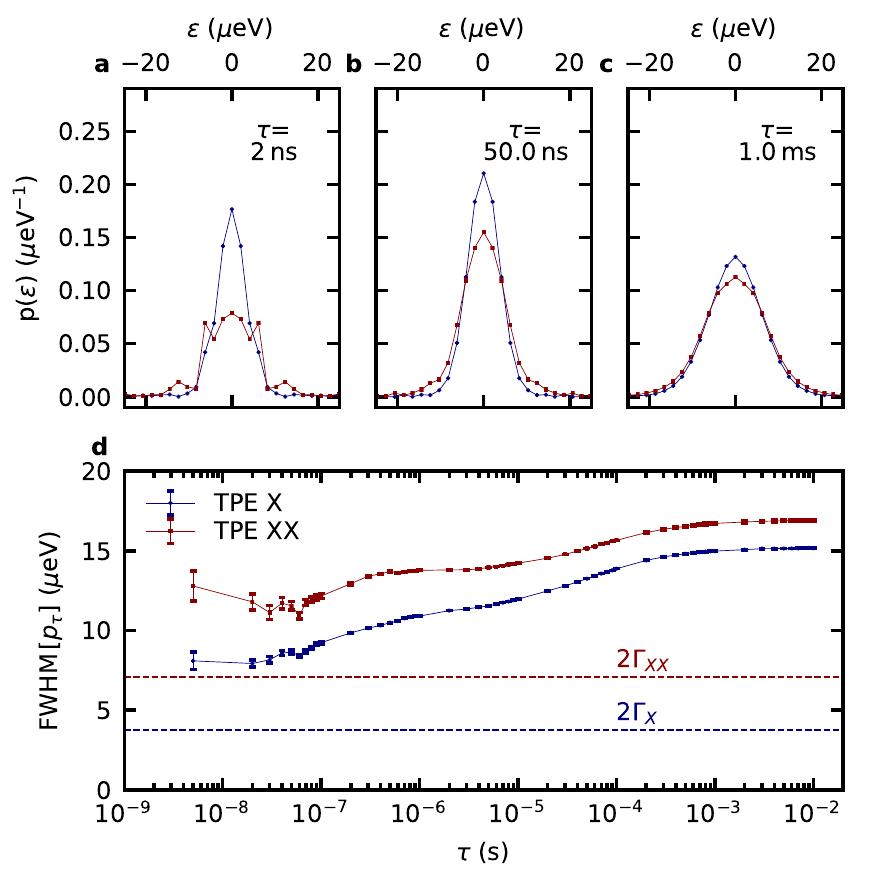}
\caption{(a-c) Distribution of spectral shifts $p(\epsilon)$ of the X and XX emission of an In(Ga)As QD at time delays $\tau$.  One FSC was selected by a polarizer. (d) Full width at half maximum (FWHM) of $p(\epsilon)$ as a function of $\tau$ in the range from \SI{2}{ns} to \SI{1}{ms}. The dashed lines correspond to the X and XX Fourier limits of $2\Gamma_{\text{X}}=\SI{3.75(6)}{\micro eV}$ and $2\Gamma_{\text{XX}}=\SI{7.09(20)}{\micro eV}$, respectively.}
\label{fig:XXevoInAs}
\end{figure}

We focus first on the temporal evolution of the line broadening of a GaAs QD displaying a relatively broad emission linewidth ($\sim\SI{60}{\micro eV}$) in time-averaged µPL.

Figures \ref{fig:XXevo} (a-c) present $p(\epsilon)$ for the X (magenta) and XX (blue) emission at three relevant time regimes. The corresponding FWHM of $p(\epsilon)$ for all evaluated values of $\tau$ is shown in figure~\ref{fig:XXevo}~(d). The dashed lines in the inset represent the X and and XX Fourier limits $2\Gamma_{\text{X}}=\SI{4.93(25)}{\micro eV}$ and $2\Gamma_{\text{XX}}=\SI{11.45(40)}{\micro eV}$, calculated from the measured lifetimes (see figure \ref{fig:spectra} (b)). Note that the actual Fourier limits can slightly vary among different QDs. At the lowest accessible time delay $\tau_{min}=\SI{2}{ns}$, X (XX) emission exhibits a FWHM (fitted by a Pseudo-Voigt function) of \SI{7.9(6)}{\micro eV} and \SI{12.6(17)}{\micro eV}. Because of the substantially different lifetimes of XX and X lines, it is not surprising to see that the XX line is broader than the X line. We attribute the discrepancy between the measured values and the respective Fourier limits to phonon coupling \cite{Reigue2017}\cite{Iles-Smith2017} and possibly charge noise.

Starting at around $\tau=\SI{10}{ns}$, a monotonically increasing broadening of $p(\epsilon)$ for both the XX and the X emission becomes apparent (due to decreasing statistical error, the increase becomes significant from approximately $\SI{100}{ns}$). 
While for short time delays the X line remains narrower than the XX counterpart, a crossover of the linewidths is observed at $\tau \approx \SI{1}{\micro s}$ (see figure \ref{fig:XXevo} (d)). All QDs measured so far showed the same behaviour. Since it is known that the XX emission is less sensitive to electric fields than the X emission \cite{Trotta2013}, the observation is consistent with a broadening stemming from fluctuating electric fields (e.g. charge noise)\cite{Kuhlmann2013}.

Finally, at the millisecond timescale both FWHM converge to $\SI{60.5(2)}{\micro eV}$ (for X) and $\SI{32.7(1)}{\micro eV}$ (for XX), which correspond to the inhomogeneously broadened spectral lines observed by a spectrometer or by Michelson interferometry.

As mentioned before, line broadening beyond the Fourier limit affects the interference visibility in a Hong-Ou-Mandel (HOM) \cite{Mandel1987} measurement between two consecutively emitted photons. For the X line and a time delay of \SI{2}{ns} we find an interference visibility of $V_{\text{HOM}}^{\text{X}}=\SI{69(3)}{\percent}$ under TPE (see supplementary for details). The visibility for the XX photons were not determined for this QD, but numerous previous measurements exhibit values similar to the X photons \cite{Huber2016}. In the XX-X decay cascade, we expect the visibility of both the XX and the X photons to be inherently limited by time-correlation between XX and X photons \cite{Simon2005}\cite{Huber2013}\cite{Troiani2014} to $V_{\text{HOM}}^{\text{max}}= \Gamma_{\text{XX}} / (  \Gamma_{\text{X}} +  \Gamma_{\text{XX}} ) = \SI{70(2)}{\percent}$. In contrast to that, resonant excitation schemes directly addressing the exciton state yield HOM visibilities above $\SI{90}{\percent}$ on the very same QDs \cite{Reindl2018Arxiv}.  The measured HOM visibility for X photons with $\tau=\SI{12.5}{ns}$ is $\SI{43(3)}{\percent}$. The decreasing HOM visibility with increasing photon delay, which is reproducible also for other QDs, is consistent with the increasing line broadening measured by PCFS and is in accordance with the work of other groups \cite{Thoma2016}.

We stress that a direct measurement of the HOM visibility by employing delay lines with variable length \cite{Loredo2016} \cite{Wang2016} certainly is the most precise approach, but likewise impractical and time consuming if intended for frequent characterization. PCFS inherently includes the full range of time delays in the measured raw data, which can be post-processed at will. Furthermore, PCFS is robust against signal intensity fluctuations, stray-light, imperfections of the transmission/reflection ratio of the BS and the spatial overlap of the beam (if constant over varying $\delta$), as all of these factors only lower the signal-to-noise ratio (SNR), but do not falsify the evaluated energy distribution. The implementation of the optical components is straightforward compared to a HOM setup, which demands high precision in temporal and spatial mode overlap.

We want to point out that - for this specific QD - the results shown in Figure~\ref{fig:XXevo}~(d) clearly show that XX is more suitable than X for HOM measurements among remote QDs \cite{Reindl2017}, such as those necessary for a Bell-state measurement used in entanglement swapping. In this case, spectral broadenings for separate emitters are uncorrelated and the deviation from the Fourier limit at long time delays is most important \cite{Weber2019} \cite{Weber2018}. 

\bigskip

To demonstrate the general applicability of PCFS, the study was repeated using an In(Ga)As QD. Figures \ref{fig:XXevoInAs} (a)-(c) show $p(\epsilon)$ for the X and XX transition for three distinct $\tau$, while \ref{fig:XXevoInAs} (d) depicts the evolution of the FWHM of $p(\epsilon)$ for $\tau$ ranging from $\SI{2}{ns}$ to $\SI{1}{ms}$. The dashed lines indicate Fourier limits of $2\Gamma_{\text{X}}=\SI{3.75(6)}{\micro eV}$ and $2\Gamma_{\text{XX}}=\SI{7.09(20)}{\micro eV}$, calculated from the respective radiative lifetimes. At a time delay of $\tau=\SI{2}{ns}$ the FWHM for the X and XX result in $\SI{8.0(5)}{\micro eV}$ and $\SI{12.8(8)}{\micro eV}$, respectively. Consecutively emitted photons in this time regime exhibit interference visibility of $V_{\text{HOM}}^{\text{X}}=\SI{57(4)}{\percent}$ and $V_{\text{HOM}}^{\text{XX}}=\SI{67(5)}{\percent}$, which is within the error of the limit of $V_{\text{HOM}}^{\text{max}}=\SI{65(2)}{\percent}$, determined by the lifetime ratio of XX and X.\\
Towards higher $\tau$, we observe again an unequal evolution of the line broadening for X and XX, due to the weaker coupling of the XX state to charge noise. At a time delay of about $\SI{10}{ms}$ the widths converge to $\SI{15.17(2)}{\micro eV}$ (X) and $\SI{16.89(7)}{\micro eV}$ (XX). In contrast to the case of the measured GaAs, the curves never cross. Furthermore they reach saturation at longer time delays, which indicates the presence of low frequency components in the charge noise.

\subsection{Influence of white light illumination on the charge dynamics of GaAs QDs}
\label{sec:white}

In the context of QDs embedded in a semiconductor host structure, additional weak white light illumination is well known to potentially enhance their brightness, especially under resonant excitation and in absence of intentional doping \cite{Michler2000, Jahn2015}. This effect is attributed to a reduction of the "blinking", which is an on-off-modulation (telegraph noise) of the QD emission, presumably due to diffusion of charges from QD into the vicinity or vice versa. Although the impact of white light on the steady-state emission conditions are well explored, little is known about actual variations in the charge dynamics.

\begin{figure}[htbp]
\centering\includegraphics[width=90mm]{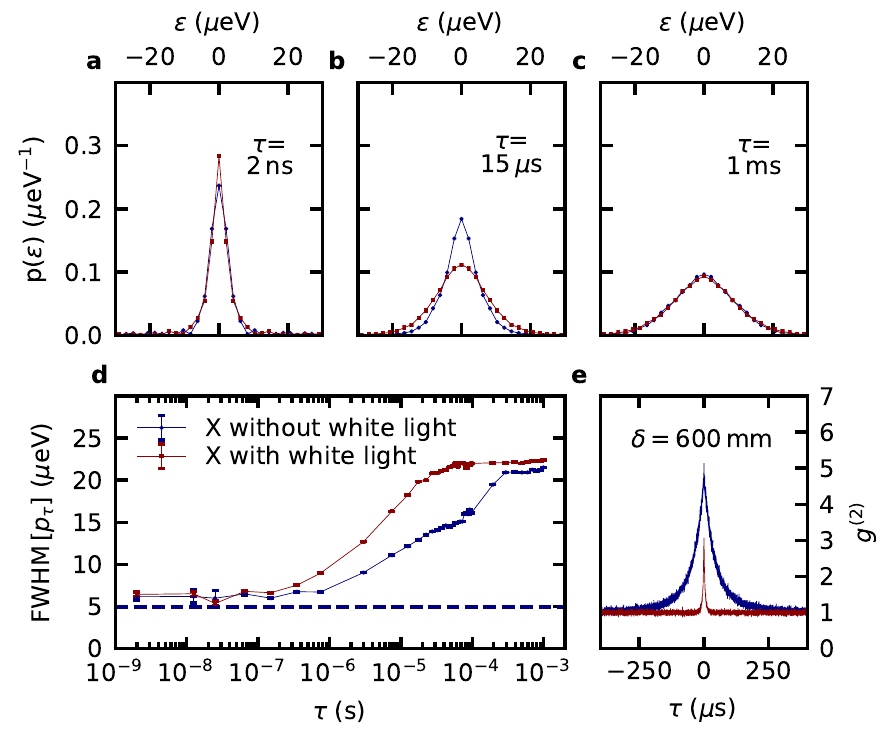}
\caption{(a-c) Distribution of spectral shifts $p(\epsilon)$ of the X signal of a GaAs QD without (blue) and with (magenta) additional white light illumination at different $\tau$. (d) FWHM of $p(\epsilon)$ as a function of $\tau$ in a range from \SI{2}{ns} to \SI{1}{ms}. The dashed line indicates the X Fourier limit of $2\Gamma_{\text{X}}=\SI{4.93(25)}{\micro eV}$. (e) Second order correlation function $g^{(2)}(\tau)$ at $\delta=\SI{600}{mm}$. The bunching, indicating the blinking of the QD emission, is reduced by white light illumination.}
\label{fig:PCFSLight}
\end{figure}

Figure \ref{fig:PCFSLight} (a-c) show the distribution $p(\epsilon)$ of the X emission from a GaAs QD, for three different $\tau$, without and with additional white light. The intensity of the white light was adjusted in order to achieve a maximum enhancement of the brightness. At short time scales (see figure~\ref{fig:PCFSLight} (a)), the lineshape is close to a Lorentzian and the FWHM of $p(\epsilon)$ remains similar for both cases, as seen in figure \ref{fig:PCFSLight} (d). Accordingly, the HOM visibility at this timescale is largely unaffected by the white light, albeit a slight trend towards higher visibility (a few percent) was observed by us on similar QDs.
The increase in line broadening discussed in the previous section is however different in the two cases: white light clearly accelerates the linewidth deterioration and substantial differences are observed at timescales of about $\SI{1}{\micro s}$.
This observation indicates that white light accelerates fluctuations of the charge environment, leading to a more significant line broadening starting at lower values of $\tau$. This is clearly visible by inspecting the lineshape with and without white light (see figure~\ref{fig:PCFSLight} (b)). When fitting with a Pseudo-Voigt function, the line shape of $p(\epsilon)$ exhibits a more Lorentzian character in the absence of white light, while a more Gaussian character is apparent when using white light.

A previous study reported a similar transition from a Lorentzian to a Gaussian broadening of the X emission line with increasing CW above-band laser pumping power (using In(Ga)As QDs)~\cite{Berthelot2006}. The effect was explained by "unconventional" motional narrowing, since the behaviour is opposite from the expected one in bulk semiconductors or quantum wells. Another study reported a different situation, where the emission linewidth from a resonantly excited GaAs QD becomes narrower with higher additional CW laser power \cite{Azzano2018}. We believe that further research, exploiting the time resolution of PCFS, will help to shine light on the connections between charge dynamics and motional narrowing.

Eventually, at large $\tau$ in the millisecond range, the FWHM of $p(\epsilon)$ converge to similar value values for both cases (figure~\ref{fig:PCFSLight} (c) and (d)).

More information on the effects produced by white light illumination can be gathered by inspecting the $g^{(2)}(\tau,\delta_i)$ shown in figure \ref{fig:PCFSLight}  (e) for $\delta_i=\SI{600}{mm}$ (A large value of $\delta$ was chosen to exclude any correlation effects stemming from interference at the BS). In absence of blinking, we would expect a flat distribution with $g^{(2)}(\tau,\delta_i)=1$ (Note that the antibunching at $\tau=0$ due to the single photon emission characteristics is irrelevant due to the large histogram time bin of \SI{100}{ns}). In presence of blinking, $g^{(2)}(\tau,\delta_i)$ has a value of $1/\beta>1$ for $\tau$ close to 0 and then decays exponentially with a characteristic time constant, which we denote as correlation time. The value of $\beta$ can be interpreted as the  on-time fraction  of the QD emission~\cite{Jahn2015}. We see that white light increases $\beta$ from about $\SI{19}{\percent}$ to $\SI{33}{\percent}$, which is in good agreement with the observed increase in brightness. Simultaneously, white light reduces the correlation time from \SI{46.7(14)}{\micro s} to \SI{4.7(4)}{\micro s}, corresponding to a faster on/off modulation of the emission. By comparing the extracted correlation time with the time at which the line broadening starts to increase in  figure~\ref{fig:PCFSLight} (d) we conclude that the increased intensity produced by white light comes at the expense of an increased charge noise at intermediate timescales $\SI{500}{ns}\lesssim\tau\lesssim\SI{500}{\micro s}$.

\section{Conclusions}
We demonstrated the implementation of a time resolved Fourier spectroscopy technique - Photon Correlation Fourier Spectroscopy (PCFS) \cite{Brokmann2006} - to study the optical emission dynamics of GaAs and In(Ga)As quantum dots under resonant two-photon excitation. We thoroughly explained the experimental implementation and interpretation of obtained data. The experiments gave access to the time evolution of the emission line broadening of GaAs and In(Ga)As quantum dots. The acquired information allow an estimation of the shape and width of the spectral distribution at different time scales, supporting a better understanding of the underlying broadening mechanisms acting in the solid state environment, such as electron-phonon coupling \cite{Reigue2017} and charge noise \cite{Kuhlmann2013}. The charge-induced inhomogeneous broadening lowers the HOM interference visibility \cite{Legero2006}, thus the probing of the temporal behaviour of the fluctuating charges allows us to deduce a trend for the indistinguishability of photons emitted at different time delays. Moreover, we investigated the effect of additional weak white light illumination on the emission of GaAs QDs and observed an acceleration of the charge dynamics, while enhancing the brightness due to suppressed blinking \cite{Michler2000, Jahn2015}.

The method stands out for its robustness against imperfection of the optical components and the tremendous information content obtained per measurement. The high achievable energy- and time resolution make PCFS a viable comprehensive characterisation method for various photonic structures, outperforming established techniques such as Michelson interferometry \cite{Santori2002, Berthelot2006}, standard diffraction spectroscopy, or even Fabry-Perot interferometry (due to the additional time resolution). The measurement is independent on the excitation conditions and can thus be performed \textit{in operando}, allowing for a reliable characterization of the emitted light as it is needed in further experiments and applications. Because the full time scale is covered by one measurement and the excitation conditions remain unaltered, PCFS is clearly a viable alternative to direct HOM measurements with varying delay lines \cite{Loredo2016} \cite{Wang2016} or scanning resonance fluorescence \cite{Kuhlmann2015} experiments. While resonant optical excitation was used here, the technique is perfectly suited for characterizing and possibly optimizing the performance of practical electrically-driven sources of quantum light \cite{Heindel2010, Salter2010, Chung2016, Zhang2015}.

\section*{Funding}
This work was financially supported by the European Research Council (ERC) under the European Unions Horizon 2020 research and innovation programme (SPQRel, Grant agreement No. 679183), the Austrian Science Fund (FWF): P 29603, the Linz Institute of Technology (LIT), the LIT Secure and Correct Systems Lab, the European Union Seventh Framework Programme (FP7/2007-2013) under grant agreement no. 601126 (HANAS), the Central European Institute of Technology (CEITEC) mobility programme (7AMB17AT044), the MEYS, the European Union's Horizon 2020 (2014-2020) research and innovation framework programme under grant agreement No731473, the project EMPIR 17FUN06 SIQUST, the Austrian Federal Ministry of Science, Research and Economy under the OEAD project CZ 07/2017, the state of Bavaria. This work has been supported by the the Deutsche Forschungsgemeinschaft (DFG) under the project SCHN1376 5.1 and the QuantERA HYPER-U-P-S project. Project HYPER-U-P-S has received funding from the QuantERA ERA-NETCofund in Quantum Technologies implemented within the European Union’s Horizon 2020 Programme.

\section*{Acknowledgments}
We thank H. Huang, K.D. Jöns, F. Basso Basset and M. Rota for fruitful discussions.

\bigskip
See Supplement 1 for supporting content.


\bibliography{main}

\end{document}